\documentclass[aps,prl,reprint,groupedaddress]{revtex4-1}

\usepackage{graphicx}
\usepackage{longtable}

\newcommand{\qt}{\tilde q}
\newcommand{\gt}{\tilde g}
\newcommand{\ul}{\tilde u_L}

\newcommand{\ur}{\tilde u_R}

\newcommand{\xc}{\tilde\chi^0_1}

\newcommand{\xcpm}{\tilde \chi^\pm_1}

\newcommand{\xt}{\tilde t_1}
\newcommand{\xtt}{\tilde t_2}
\newcommand{\xb}{\tilde b_1}
\newcommand{\xbb}{\tilde b_2}

\begin{document}


\title{Top jets as a probe of degenerate stop-NLSP LSP scenario in the framework of cMSSM}


\author{\vskip -10pt Kirtiman Ghosh$^1$, Katri Huitu$^2$, Jari Laamanen$^3$ and Lasse Leinonen$^4$}
\affiliation{$^{1,2,4}$Department of Physics, and Helsinki Institute of Physics, FIN-00014 University of Helsinki, Finland}
\affiliation{$^3$Theoretical High Energy Physics, IMAPP, Faculty of Science, Radboud University Nijmegen Mailbox 79, P.O. Box 9010, NL-6500 GL Nijmegen, The Netherlands\\
$^{1}$kirti.gh@gmail.com, $^{2}$katri.huitu@helsinki.fi, $^{3}$j.laamanen@science.ru.nl, $^{4}$lasse.leinonen@helsinki.fi}




\begin{abstract}
The degenerate stop NLSP and neutralino LSP scenario is well motivated but hard to detect at the collider experiments. We propose a novel signature for detection this scenario at the Large Hadron Collider (LHC) and demonstrate its feasibility.  
\end{abstract}
\pacs{}
\maketitle


%

Supersymmetry (SUSY) is one of the most attractive extensions beyond the standard model (SM) and search for SUSY is one of the prime goals of the Large Hadron Collider (LHC). 
The production of colored SUSY particles, such as squarks and gluinos, at the LHC plays the most important role in discovering SUSY. SUSY signature should show up best in multi-jets plus missing transverse energy ($E_T\!\!\!\!\!\!/~$) channel in $R$-parity conserving scenarios. $E_T\!\!\!\!\!\!/~$ results from the weakly interacting stable lightest SUSY particles (LSPs) appearing in the final states. 
With 4.7 (4.98) fb$^{-1}$ integrated luminosity ($\int \!\!{\cal L} dt$) at 7 TeV center-of-mass energy, no SUSY signature has been detected in the  multi-jets plus $E_T\!\!\!\!\!\!/~$ channel by the ATLAS \cite{atlas} (CMS \cite{cms}) collaboration.  
As a result, limits are imposed on the squark and gluino masses in the context of constrained minimal supersymmetric extension of the Standard Model (cMSSM) \cite{cmssm}. The phenomenology of cMSSM is determined in terms of five parameters, namely, common scalar $m_0$ and gaugino $m_{1/2}$ mass, tri-linear mass parameter $A_0$, ratio of Higgs vacuum expectation values ${\rm tan}\beta$ and sign of Higgs mixing parameter ${\rm sign}(\mu)$. In the context of cMSSM with ${\rm tan}\beta=10$, $A_0=0$ and $\mu>0$, for large $m_0$, gluino mass below about 900 GeV is excluded at the 95\% confidence level (C.L.) from ATLAS and CMS multi-jets plus $E_T\!\!\!\!\!\!/~$ scarch. Equal mass squarks and gluinos are excluded below 1400 GeV.

The main motivation for postulating the existence of SUSY is to stabilize the electroweak hierarchy against radiative corrections. In view of this motivation, the lower bounds on the squarks and gluino masses seems already rather large. However, in the radiative corrections to Higgs boson mass parameters, third generation (s)quarks are most important at one loop order. Moreover, the ATLAS and CMS searches are not sensitive to the direct pair production of only third generation squarks \cite{3rdsusearch}. 
Therefore, stop masses of a few hundred GeV are still allowed and, in fact, favored by finetuning arguments.

The stop can be the next-to-lightest supersymmetry particle (NLSP). This can be realized in the context of cMSSM because of large third-generation Yukawa coupling. The off-diagonal trilinear term is proportional to Yukawa couplings. Therefore, it can generate large mixing and hence, large mass splitting in the stop sector and consequently produce a light stop ($\xt$) as NLSP.


There are several phenomenological motivations to be interested in the $\xt$-NLSP. The $\xt$-NLSP scenario is well-motivated to explain the relic density (RD) measured by WMAP \cite{wmap}. In the cMSSM, the lightest neutralino ($\xc$) is mostly bino–like and thus, gives rise to RD which is too large. The observed relic density can be explained if the $\xt$-NLSP coannihilate with the bino–like $\xc$-LSP. Furthermore, the electroweak (EW) baryogenesis favors a light-$\xt$ \cite{EWB}. In fact,  a SUSY scenario with a bino–like $\xc$-LSP and a $\xt$-NLSP can simultaneously explain the observed baryon asymmetry in the universe as well as the RD. Finally, cMSSM with $\xt$-NLSP is also favored in view of a Higgs boson mass ($m_h$) about 125 GeV observed by the LHC \cite{higgs}. In order to get a Higgs boson around 125 GeV, we must have either significant stop mixing or a large stop mass scale. Large stop mixing in cMSSM with $\xt$-NLSP may  give rise to $m_h$ around 125 GeV.


Collider searches for $\xt$-NLSP at the LHC is straightforward as long as $m_{\xt}- m_{\xc}$ 
is sufficiently large. In this case, $\xt$ signature should pop up in multi-jets plus $E_T\!\!\!\!\!\!/~$ search. If the $\xt$ and $\xc$ mass splitting is small (which is favored by RD) then the decay channels $\xt \to bW \xc$ and $t\xc$ are kinematically forbidden and the four-body decay $\xt \to f \bar f^{\prime} b \xc$ is strongly phase space suppressed. The dominant decay mode will be loop induced two-body decay $\xt \to c \xc$. However, due to small $\xt$-$\xc$ mass splitting, the charm jets as well as $E_T\!\!\!\!\!\!/~$ will become very soft at the collider. Therefore, di-jet $+$ $E_T\!\!\!\!\!\!/~$ signature resulting from $\xt\xt$ production will be buried deeply in the background events and cannot been selected out with the current LHC search approaches \cite{stop_buried}. Several alternative solutions to detect signatures of light $\xt$ nearly degenerate with the $\xc$ have been proposed. $\xt$ pair production in association with two $b$-jets \cite{b_stop}, single photon \cite{photon_stop} or jet \cite{jet_stop} have been studied. Stop pair production with a hard jet is the most promising channel and can probe $m_{\xt}$ upto $250$ GeV \cite{jet_stop} at the LHC with $\sqrt s=14$ TeV and $\int \!\!{\cal L} dt=$100 fb$^{-1}$.

In this letter, we propose a novel collider signature for light stop search. In cMSSM with $\xt$-NLSP, gluino ($\gt$) is heavier than $\xt$ by a factor $\sim$ 5 and thus, at the LHC, $\gt\gt$ production cross-section is atleast two orders of magnitude smaller than the $\xt\xt$ production cross-section. However, we find that $\gt\gt$ production gives rise to a interesting signature which could be observed with a few fb$^{-1}$ luminosity of the LHC running at $\sqrt s=14$ TeV.  

Gluino is lighter than the first two generation squarks ($\qt$) in most part of the cMSSM parameter space. If $\xt$ is the NLSP then $\gt$ will dominantly decay to the $t \xt$ pairs followed by  $\xt\to c\xc$ decay. In this case, $\gt\gt$ production gives rise to two top quarks, two charm quarks and $E_T\!\!\!\!\!\!/~$: 
$
pp \to \gt \gt \to t t ~\xt^* \xt^* \to tt ~\bar c \bar c~\xc\xc.
$
Since $m_{\gt}\gg m_{\xt}$, the top and the stop arising from the gluino decay will be highly boosted. However, the charm quarks resulting from the decay of the boosted stops will be soft because of their small mass and small mass difference between stops and LSPs. The momentum of the boosted stops will be carried away by the massive lightest neutralinos. Top quarks further decay hadronically ($t\to b q \bar q^{\prime}$) or leptonically ($t \to b l \nu$). However, the highly boosted tops will decay to collimated collections of particles that look like single jets. It has been already established that top jets can now be considered as standard objects for event analysis at the LHC \cite{ATL-COM-PHYS-2008-001,top_jet}. In this letter. we have studied two high transverse momentum ($p_T$) top jets  plus large $E_T\!\!\!\!\!\!/~$ as a signature of cMSSM with $\xt$-NLSP.

Gluino pair production in this scenario also gives rise to usual multi-jet$+E_T\!\!\!\!\!\!/~$ (when tops decay hadronically), opposite-sign dilepton+$E_T\!\!\!\!\!\!/~$ (when $\gt\gt$ decays to $t\bar t$ and tops decay leptonically) and same-sign dilepton (SSD)$+E_T\!\!\!\!\!\!/~$ \cite{ssdl} (when $\gt\gt$ decays to same sign tops and tops decay leptonically) signature. However, all these signatures suffer from the SM $t \bar t$ and QCD multi-jet backgrounds. Two high $p_T$ top-tagged jets+$E_T\!\!\!\!\!\!/~$ signature has the following advantages:(i) The mis-tagging efficiency of gluon or light quark jets to be tagged as top jets is very small~\cite{top_jet}. Therefore, top tagging significantly reduces SM multi-jet background.
(ii) Large $p_T$ cuts could be applied on the top-tagged jets to reduce the SM $t\bar t$ background.
(iii) Finally, this is a characteristic signature of cMSSM with $\xt$-NLSP which is well motivated from WMAP data, EW baryogenesis and the LHC indication of $m_h \sim 125$ GeV.

In Ref.~\cite{ATLAS_tt}, ATLAS collaboration has presented a search for new phenomena in $t\bar t+E_T\!\!\!\!\!\!/~$ events at the LHC with $\sqrt s=7$ TeV and $\int \!\!{\cal L} dt=$1.04 fb$^{-1}$. 
In view of consistency of the data and SM expectations, 95\% C.L. upper limit of $1.1$ pb is set on cross-section times branching ratio of any new physics process which gives rise to $t\bar t+E_T\!\!\!\!\!\!/~$ signature. Similar signature results from $\gt \gt$ production in cMSSM with $\xt$-NLSP. Assuming 100\% branching fraction for $\gt \to t \xt^*$, naive estimation of next-to-leading order (NLO) $\gt\gt$ production cross-section implies an lower bound about $m_{\gt} \sim$ 560 GeV in this scenario.

 \begin{figure}
\vskip -23pt
 \includegraphics[angle=-90,width=80mm]{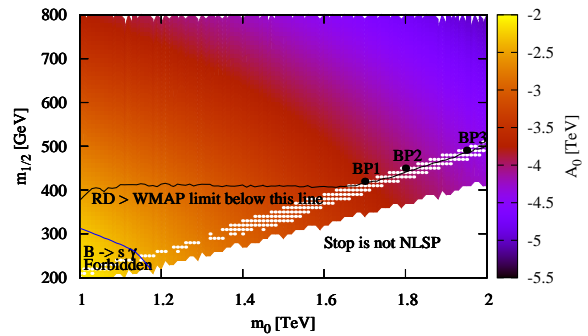}
\vskip -15pt
 \caption{The 3-dimensional parameter space gives rise to stop NLSP, for ${\rm tan}\beta=15$ and $\mu>0$. To select a particular value for $A_0$, we demand $m_{\xt}-m_{\xc} \sim 35$ GeV. The white dots corresponds to the points resulting into a $m_h\sim 125\pm 2$ GeV.}
\label{param_space}
\vskip -2pt
 \end{figure}

In cMSSM, $\xt$ can be the NLSP when the mixing term of the left and right stop states, $M_{LR}^{2}=m_{t}(A_t-\mu{\rm cot}\beta)$, is large. The largest mixing occurs when $\mu$ and $A_t$ have opposite signs. In our analysis, we consider $\mu>0$ and thus, scan over negative values of $A_0$. In addition to the mixing, the renormalization group running plays an important role in the determination of the NLSP. In the scalar RGEs, the terms proportional to the Yukawa couplings (and scalar masses) decrease the soft masses, while the terms proportional to the gauge couplings (and gaugino masses) have an opposite effect. Due to the largeness of the strong coupling constant, squark masses tend to increase more than the slepton masses, even though the third-generation squarks have large Yukawa couplings. Thus, a small $m_{1/2}$ compared to $m_0$ is preferred in order to suppress the strong coupling term in the squark RGE, which then may lead to a stop NLSP instead of stau NLSP. In our parameter space scan, we have varied $m_0$ over 1 to 2 TeV and $m_{1/2}$ over 200 to 800 GeV. Fig.~\ref{param_space} shows a 3-dimensional scan over $m_0$ (along $x$-axis), $m_{1/2}$ (along $y$-axis) and $A_0$ (color gradient) for ${\rm tan}\beta=15$ and $\mu>0$. In this parameter space, $\xt$ is always NLSP. For a given $m_0$ and $m_{1/2}$, there is a range of $A_0$ values which give rise to $\xt$-NLSP. In Fig.~\ref{param_space}, we consider the particular $A_0$ (motivated form the observed WMAP RD \cite{huitu}) for which $m_{\xt}-m_{\xc} \sim 35$ GeV. The particle spectrum was calculated using {\bf SuSpect} \cite{suspect} with $m_t=173.2$ GeV, $m_b=4.2$ GeV, $m_{\tau}=1.777$ GeV and $\alpha_s(m_Z)=0.1172$. 
 \begin{table}
\vskip -15pt
 \caption{Benchmark parameters in GeV (${\rm tan}\beta=15$, $\mu>0$).}
\label{BPs}
 \begin{ruledtabular}
\begin{tabular}{||c|c|c||c|c|c||c|c|c||}
\multicolumn{3}{||c||}{BP1} & \multicolumn{3}{c||}{BP2} & \multicolumn{3}{c||}{BP3}\\\hline
$m_0$ & $m_{1/2}$ & $A_0$ &$m_0$ & $m_{1/2}$ & $A_0$ & $m_0$ & $m_{1/2}$ & $A_0$ \\\hline\hline
1700 & 420 & -3763 & 1800 & 450 & -4001 & 1950 & 491 & -4349 \\

 \end{tabular}
 \end{ruledtabular}
\vskip -15pt
 \end{table}

For this parameter space, we have checked the following constraints:
For {\em $b\to s \gamma$ branching ratio}, we consider: $BR(B\to X_s \gamma)=(355\pm 142)\times 10^{-6}$, including the theoretical uncertainty in the SM as well as in the MSSM \cite{bsg}. We have also checked that {\em $B_s\to \mu^+ \mu^-$ branching ratio} is smaller than the experimental upper limit \cite{bmm}: $BR(B_s\to \mu^+ \mu^-) < 4.3 \times 10^{-8}$ (95\% C.L.). The new physics (NP) contribution to {\em the $BR(B^+\to\tau^+\nu_{\tau})$} can be quantified by defining: $R^{NP}_{\tau \nu_{\tau}}=\frac{BR(B^+\to\tau^+\nu_{\tau})_{SM+NP}}{BR(B^+\to\tau^+\nu_{\tau})_{SM}}$. The 95\% C.L. allowed range for $R^{NP}_{\tau \nu_{\tau}}$ is $0.99<R^{NP}_{\tau \nu_{\tau}}<3.19$ \cite{btn}. With the combined data from WMAP, BAO (Baryon Acoustic Oscillations) in the distribution of galaxies and observation of Hubble constant, the density of cold dark matter in the
universe is determined to be \cite{wmap} $\Omega_ch^2=0.1126\pm0.0036$. We consider the preferred RD range of $0.0941<\Omega_ch^2<0.1311$ at $2\sigma$ level. We have used {\bf micrOmegas} (v.2.4.R) \cite{microomega} for calculating RD and other constraints. Fig.~\ref{param_space} shows that large part of the parameter space is below
the WMAP upper limit (and can be made consistent with the preferred RD range)
as well as consistent with other experimental observations.


 \begin{table}
\vskip -15pt
 \caption{Masses (in GeV) of relevant sparticles.}
\label{mass}
 \begin{ruledtabular}
 \begin{tabular}{||c||c|c|c|c|c|c||c|c|c||}
  &  $\gt$ & $\ul$ & $\ur$ & $\xtt$ & $\xt$ & $\xbb$ & $\xb$ & $\xcpm$ & $\xc$ \\\hline\hline
BP1 & 1064 & 1880 & 1872 & 1334 & 217 & 1776 & 1287 & 352 & 181 \\
BP2 & 1133 & 1994 & 1985 & 1411 & 229 & 1883 & 1365 & 378 & 194 \\
BP3 & 1227 & 2161 & 2150 & 1523 & 250 & 2040 & 1478 & 414 & 213 \\
\end{tabular}
 \end{ruledtabular}
 \end{table}
The ATLAS and CMS collaborations \cite{higgs} have presented the first indication for a Higgs boson with $m_h \sim 125\pm 2$ GeV. 
Due to the large stop mixing, the $\xt$-NLSP scenario can accommodate such a Higgs boson mass. In Fig.~\ref{param_space}, the white dots correspond to the points resulting into a $m_h \sim 125\pm 2$ GeV. Fig.~\ref{param_space} shows that small part of $m_0$-$m_{1/2}$ plane is consistent with both $m_h \sim 125\pm 2$ GeV and WMAP data as well as other constraints. With this motivation, we have chosen three benchmark points (BP), listed in Table ~\ref{BPs}, for presenting our numerical results. The masses of relevant sparticles for the BPs are presented in Table~\ref{mass}. In Table~\ref{cons}, we present the Higgs boson mass, RD and other constraints for the three BPs. 

 \begin{table}
\vskip -15pt
 \caption{Model contribution to different observables.}
\label{cons}
 \begin{ruledtabular}
 \begin{tabular}{||c||c|c|c|c|c||}
BPs & $m_h$ & $\Omega_c h^2$ & $\rm{BR}(B \to X_s \gamma)$ & $\rm{BR}(B_s \to \mu^+\mu^-)$ & $R^{\rm NP}_{\tau \nu_\tau}$ \\\hline\hline
BP1 & 124.0 & 0.115 & $2.77\times 10^{-4}$ & $3.14\times 10^{-9}$ & 0.997  \\
BP2 & 124.1 & 0.108 & $2.82\times 10^{-4}$ & $3.13\times 10^{-9}$ & 0.998  \\
BP3 & 125.3 & 0.131 & $2.89\times 10^{-4}$ & $3.12\times 10^{-9}$ & 0.998  \\
 \end{tabular}
 \end{ruledtabular}
\vskip -15pt
 \end{table}

We consider boosted top jets as the signature of cMSSM with $\xt$-NLSP. To detect top jets, we have used Johns Hopkins top tagger algorithm \cite{top_jet}. A brief description of the algorithm is as follows: Particles are clustered into jets of size $R$ using the Cambridge-Aachen (CA) algorithm \cite{ca}. To find subjets, each jet in the event is declustered into two objects by reversing the CA clustering. The softer object is thrown out if the ratio of its $p_T$ and the full jet $p_T$ is less than a parameter $\delta_p$, and the declustering continues on the harder object. The declustering stops and the original jet is considered irreducible when the $p_T$ ratios of both objects are smaller than $\delta_p$, two objects are too close $|\Delta \eta|-|\Delta \phi|<\delta_r$ or there is only one cell left. The original jet declusters into two subjets if the $p_T$ ratios of both objects is larger than $\delta_p$ and the previous steps are repeated on those subjets. To reconstruct boosted hadronic top jets, cases with 3 or 4 subjets are selected for the further analysis.

\begin{table}
\vskip -15pt
 \caption{Signal and SM background cross-sections at the LHC with $\sqrt s=14$ TeV after different set of cuts.}
\label{cs}
 \begin{ruledtabular}
 \begin{tabular}{||c||c|c|c|c||}
 & \multicolumn{4}{c||}{SM Background cross-section in fb} \\\cline{2-5}
 Process  & Acc. & Acc. Cuts $+$ & Acc. Cuts $+$ & Acc. Cuts $+$ \\
      & Cuts  & $E_T\!\!\!\!\!\!/\ge 100$ GeV& $E_T\!\!\!\!\!\!/\ge 300$ GeV& $E_T\!\!\!\!\!\!/\ge 500$ GeV\\\hline\hline
$t \bar t$   & 60.09 &    1.29   &    8.1$\times 10^{-2}$  &   $>$1.5$\times 10^{-3}$ \\
${\rm jets}$   & 51.32 &    0.26  &    6.2$\times 10^{-3}$  &   $>$3.1$\times 10^{-3}$ \\
$t \bar t Z$ & 0.0022 & 1.6$\times 10^{-3}$  & 3.2$\times 10^{-4}$  &  $>$3.4$\times 10^{-4}$ \\
$t \bar t W$ & 0.074  & 2.7$\times 10^{-2}$  & 2.8$\times 10^{-3}$  &  9.2$\times 10^{-4}$  \\\hline\hline
 BPs & \multicolumn{4}{c||}{Signal cross-section in fb} \\\hline\hline
BP1 & 2.49 & 2.32 & 1.51 & 0.76 \\
BP2 & 1.75 & 1.66 & 1.19 & 0.73 \\
BP3 & 1.22 & 1.16 & 0.87 & 0.57 \\

 \end{tabular}
 \end{ruledtabular}
\vskip -15pt
 \end{table}

In our analysis, we consider the events with total hadronic energy deposits, $E_T>1$ TeV and use {\bf FASTJET} v.3.0.2 \cite{fjet} for the clustering and subjet analysis. Since more highly boosted tops will be more collimated, following Ref.~\cite{top_jet}, we choose $R$ and $\delta$ parameters as: for $E_T>1,~1.6,~2.6$ TeV, $R=0.8,~0.6,~0.4$ and $\delta_{p(r)}=0.10(0.19),~0.05(0.19),~0.05(0.19)$, respectively. A jet is considered as top-tagged if (i) it contains 3 or 4 substructure, (ii) the invariant mass of the sum of the subjets momentum four-vectors falls in the range $150$-$200$ GeV ($\sim m_t\pm 25$ GeV) and (iii) there exist two subjets which reconstruct the W-mass to within $65$-$95$ GeV. 

We have used {\bf Prospino} 2.1 \cite{prospino} with CTEQ6.6M \cite{cteq6.6m} parton distribution function (pdf)for computing NLO $\gt\gt$, $\qt\gt$, $\qt^* \gt$, $\qt\qt$ and $\qt\qt^*$ production cross-sections at the LHC with $\sqrt s=14$ TeV. The QCD factorization and renormalization scales are kept fixed at the sum of masses of the produced SUSY particles. Signal events, showering, hadronization, initial state and final state radiation are simulated by {\bf PYTHIA} 6.421 \cite{pythia}. An event is selected for further analysis if it contains at least two top-tagged jets with $p_T \ge 500$ GeV and $|\eta| \le 2.5$. We reject events with one or more isolated electron or muon \cite{isolated} with $p_T\ge 20$ GeV and $|\eta| \le 2.5$. We collectively refer this set of cuts as {\em Acceptance Cuts}. The signal is characterized by large $E_T\!\!\!\!\!\!/~$ resulting from the boosted $\xc$ which escape the detector without being detected. Signal cross-sections for the BPs are presented in Table~\ref{cs} after acceptance and different $E_T\!\!\!\!\!\!/~$ cuts ($E_T\!\!\!\!\!\!/~\ge 0,~100,~300~{\rm and}~500$ GeV). 

The dominant SM background arises from the $t \bar t$ production. The efficiency for gluon or quark jets to be mistagged as top jets is very small \cite{top_jet}. However, due to huge production cross-section, QCD multi-jet events might also contribute to the background. In these contributions, $E_T\!\!\!\!\!\!/~$ arises from momentum mis-measurement due to finite detector resolution. We have approximated the detector resolution effects by smearing the jet energies with Gaussian functions \cite{smear}. Table~\ref{cs} shows that large $E_T\!\!\!\!\!\!/~$ cuts reduce $t \bar t$ and multi-jets backgrounds to a negligible level. Production of $t \bar t$ pairs with a $Z$ or $W$-boson may contribute to the background when $Z$ decays invisibly and $W$ decays leptonically. Lepton veto in the acceptance cuts significantly reduce the $t \bar t W$ contribution. Multi-jets and $t\bar t$ were simulated with {\bf PYTHIA}. Whereas, $t\bar t Z$ and $t \bar t W$ events were generated using {\bf ALPGEN} \cite{alpgen}, and the unweighted event samples were analyzed by {\bf PYTHIA}. For background analysis, we have used CTEQ6L1 \cite{cteq6l1} pdf. The renormalization and factorization scales have been set equal to the subprocess center-of-mass energy $\sqrt {\hat s}$. For $t \bar t$ production, we have used NLO+NNLL cross-section of 877 pb at 14 TeV LHC. In Table~\ref{cs}, we present SM background contributions after different $E_T\!\!\!\!\!\!/~$ cuts. 

 \begin{figure}
\vskip -20pt
 \includegraphics[angle=-90,width=80mm]{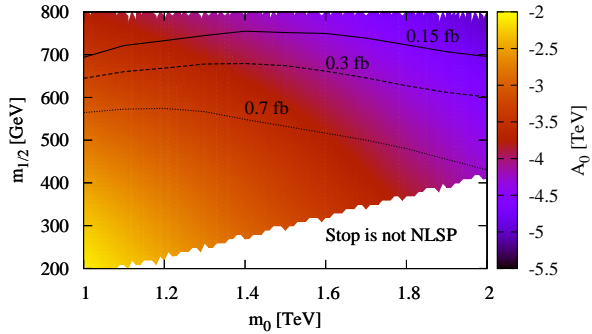}
\vskip -15pt
 \caption{Constant signal cross-section contours in $m_0$-$m_{1/2}$ plane after Acc. cuts + $E_T\!\!\!\!\!\!/~\ge 500$ GeV at the 14 TeV LHC.}
\vskip -15pt
\label{ccc}
 \end{figure}

\begin{figure}[t]
 \includegraphics[angle=-90,width=80mm]{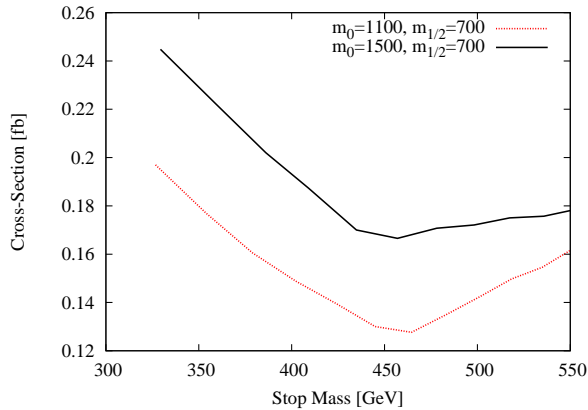}
\vskip -8pt
 \caption{Signal cross-sections as a function of stop mass for two different set of $m_0$-$m_{1/2}$ after  Acc. cuts + $E_T\!\!\!\!\!\!/~\ge 500$ GeV at the 14 TeV LHC. ${\rm tan}\beta=15$ and $\mu>0$ are assumed.}
\label{stop_cross}
\vskip -15pt
 \end{figure}

According to Table~\ref{cs}, for  $E_T\!\!\!\!\!\!/~>500$ GeV, the number of predicted SM background event with $\int \!\!{\cal L} dt=100$ fb$^{-1}$ is less than one. If we demand $5 (10)$ signal events for the discovery, BP1, BP2 and BP3 will be probed with 7(13), 7(14) and 9(18) fb$^{-1}$ integrated luminosity of the LHC running at $\sqrt s=14$ TeV. In Fig.~\ref{ccc}, we present constant signal cross-section contours on $m_0$-$m_{1/2}$ plane after {Acc. cuts + $E_T\!\!\!\!\!\!/~\ge 500$ GeV} at the LHC with $\sqrt s=14$ TeV. Assuming $5 (10)$ signal events for the discovery, 
the $0.7,~0.3~{\rm and}~0.15$ fb cross-section contours (equivalently, for large $m_0$, $m_{\gt}\sim$ 1.1, 1.5 and 1.7 TeV ) can be probed with $\int \!\!{\cal L} dt=7(13),~17(33)~{\rm and}~33(67)$ fb$^{-1}$.

Up to this point, we have discussed two high-$p_T$ top-tagged jets$+E_T\!\!\!\!\!\!/~$ as the signature of nearly degenerate $\xt$-NLSP LSP scenario. To show the feasibility of this signal for non-degenerate $\xt$-$\xc$ scenario, we have varied $A_0$ and hence, $m_{\xt}$, keeping $m_0$ and $m_{1/2}$ fixed. In Fig.~\ref{stop_cross}, we present the signal cross-sections (for 14 TeV LHC) after  Acc. cuts + $E_T\!\!\!\!\!\!/~\ge 500$ GeV as a function of $m_{\xt}$. Initially signal cross-section decreases because increasing $m_{\xt}$ implies less $\gt$-$\xt$ mass splitting and hence, less boosted tops. In Fig.~\ref{stop_cross}, $m_{1/2}=700$ GeV corresponds to $m_{\xc}\sim 300$ GeV. For $m_{\xt}>475$ GeV, boosted $\xt$ decay to $t \xc$ increases the possibility of getting more boosted top jets and thus, enhances the signal cross-section. Fig.~\ref{stop_cross} shows that signal cross-section does not vary much as long as stop is the NLSP. Therefore, the signature under consideration is equally applicable for the whole $\xt$-NLSP region of the cMSSM parameter space.

To summarize, we have investigated two high-$p_T$ top-tagged jets$+E_T\!\!\!\!\!\!/~$ as the signature of cMSSM with degenerate $\xt$-NLSP LSP. 
We have considered $\gt \gt$ production. Boosted tops results from $\gt \to t \xt$ decay. We find that as long as stop is the NLSP (including degenerate $\xt$-NLSP LSP scenario), collider searches for high-$p_T$ top jets plus large $E_T\!\!\!\!\!\!/~$ can probe gluino mass upto 1.7 TeV at the 14 TeV LHC with $33$ fb$^{-1}$ integrated luminosity.

\begin{acknowledgments}
The work of JL was supported by the Foundation for Fundamental
Research of Matter (FOM), program 104 ``Theoretical Particle Physics in
the Era of the LHC".
\end{acknowledgments}


\begin{thebibliography}{99}
\bibitem{isolated} For isolated leptons, the total energy deposit from all hadronic activity within a cone of $\Delta R \le 0.2$ around the lepton axis should be  $\le 10$ GeV.
\end{thebibliography}

\end{document}